\documentclass[twocolumn,showpacs,preprintnumbers,amsmath,amssymb]{revtex4}

\usepackage{graphicx}
\usepackage{dcolumn}
\usepackage{bm}

\begin{document}

\title{Uniqueness of Fisher plots and their inferred (nuclear) liquid-vapor phase diagrams}

\author{J. B. Elliott, L. G. Moretto and L. Phair}
\affiliation{Nuclear Science Division, Lawrence Berkeley National Laboratory,
Berkeley, CA  94720}

\date{\today}
\begin{abstract}
A detailed study of the lattice gas (Ising) model shows that the correct definition of cluster concentration in the vapor at coexistence with the liquid leads to Fisher plots that are unique, independent of fixed particle number density  $\rho_{\text{fixed}}$, and completely reliable for the characterization of the liquid-vapor phase diagram of any van der Waals systems including nuclear matter.
\end{abstract}

\preprint{LBNL-54982}

\maketitle

The liquid to vapor phase transition in nuclei and in nuclear matter is a time honored topic nearing its culmination with a complete phase diagram of bulk nuclear matter \cite{moretto-04}.\ \ On one hand, general indicators of phase transitions in finite systems have been explored theoretically \cite{jaqaman-83,goodman-84,bondorf-95,gross-97}, while, on the other, experimental evidence for phase transitions has been claimed \cite{finn-82,campi-86,gilkes-94,pochodzella-00,dagostino-00,elliott-02}.

Specifically, one method based on Fisher's droplet model \cite{fisher-67} was applied to experimental nuclear multifragmentation data and the full liquid-vapor coexistence line (phase diagram) from low temperatures up to the critical point was extracted \cite{moretto-04,elliott-02}.

The validity of such an approach has been demonstrated by applying it to the cluster concentrations generated by the Ising model \cite{mader-03}.\ \ The cluster concentrations collapse onto a single line (a Fisher plot) according to Fisher's theory \cite{fisher-67} and allow for the quantitative recovery of the Ising phase diagram and the associated parameters, e.g. the critical temperature $T_c$.

This success has been tempered by the claim that the Ising cluster concentrations, while giving excellent Fisher plots, depend upon the mean particle number densities $\rho_{\text{fixed}}$ for which they have been calculated, and project different critical points for different mean densities \cite{gulminelli-02,dagostino-03}.  The implication of these claims is that the phase diagram for the Ising model and for nuclear matter obtained by such procedures is not unique and hence incorrect.

In this Rapid Communication it will be shown that such claims arise from an improper use of Fisher's theory, and that a correct application of that theory leads to a unique Fisher plot, independent of the value of $\rho_{\text{fixed}}$, which therefore can be directly transcribed into a unique phase diagram with the correct critical parameters.

Preliminary to the present analysis of cluster concentrations of the Ising model, it is worth recalling that, for an infinite system, thermodynamics demands that the saturated vapor properties be independent of the mean density of the system.\ \ Consider a cylinder with a given amount of water and a piston that can alter the volume of the fluid.\ \ At \underline{any} fixed temperature $T$ below the critical temperature, the vapor in equilibrium with the liquid has a pressure $p(T)$ which is completely determined by the temperature (univariance).\ \ The position of the piston, to which the appropriate counter pressure $p(T)$ is applied, is in indifferent equilibrium (and the system has an infinite isothermal compressibility), i.e. it can be moved up or down without experiencing any restoring force, thus covering the entire density range from liquid ${\rho}_l$ to vapor ${\rho}_v$ until the last drop of liquid remains.\ \ The properties of the vapor in the univariant regime depend exclusively on $T$ and not on the total amount of vapor or liquid in the system because at phase equilibrium (coexistence) the chemical potential is uniquely defined by $T$.\ \ Thus the corresponding Fisher plot depends only on the property of the vapor and \underline{must} be independent of $\rho_{\text{fixed}}$.

After the last drop of liquid disappears the vapor is free from such a constraint.\ \ As the piston further rises after the disappearance of the liquid, the system becomes bivariant, and the piston is no longer at indifferent equilibrium, but must obey the equation of state: $p = f(\rho,T)$.\ \ For the ideal gas limit $p \propto \rho T$.\ \ In these conditions, deviations from the Fisher plot are to be expected unless the bivariance is accounted for in terms of the chemical potential.\ \ Thus the obvious caveat: \emph{the standard Fisher plot is unique and independent of the mean density provided only cluster concentrations at coexistence are considered}.

\begin{figure}[ht]
\includegraphics[width=8.7cm]{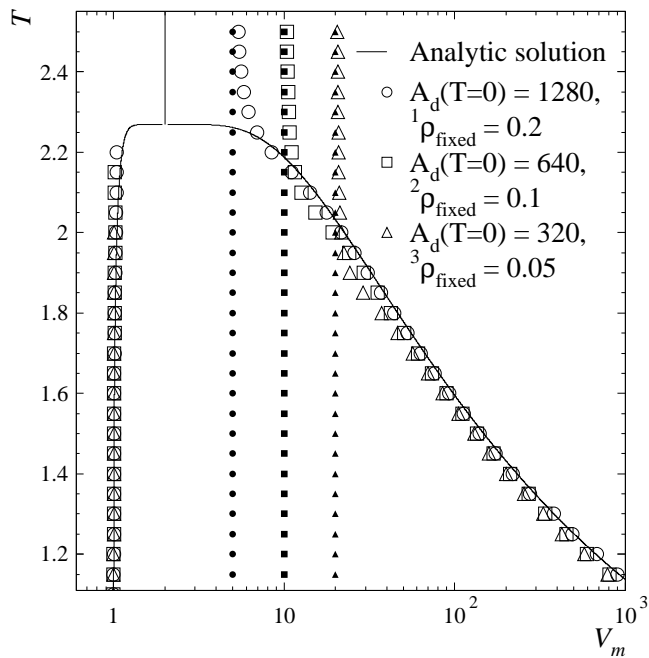}
\caption{Temperature $T$ as a function of molar volume  $V^{l,v}_m$ of liquid (left) and vapor (right) for different fixed mean densities $\rho_{\text{fixed}}$ compared to the analytic result.\ \ Deviations from the analytic result are due to finite size effects of the drop.}
\label{t-vm}
\end{figure}

\begin{figure}[ht]
\includegraphics[width=8.7cm]{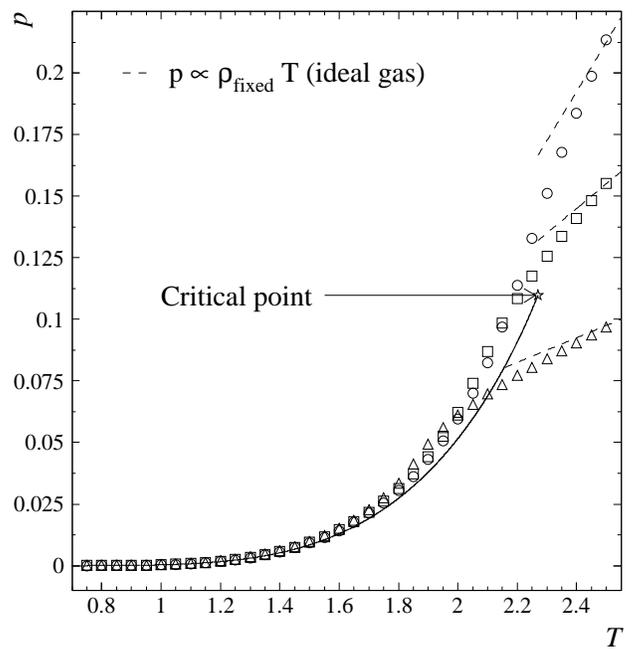}
\caption{Pressure $p$ as a function temperature $T$.\ \ The pressure stays on the coexistence curve until complete evaporation and ideal behavior is observed.\ \ Deviations from the analytic solution are due to the finite size of the drop.}
\label{p-t}
\end{figure}

The features discussed above can be observed in cluster concentrations generated from fixed magnetization ($M_{\text{fixed}}$) Ising calculations on a two dimensional ($d=2$) square lattice of side $L$.\ \ The calculations are performed via the method of Heringa and Bl\"{o}te \cite{heringa-98}.\ \ Periodic boundary conditions and a lattice side of $L=80$ are used to minimize the finite size effects of the container; a shift of only $\sim0.45$\% in the value of $T_c$ is expected for this system \cite{ferdinand-69,landau-76}.\ \ The ground state liquid drops, $A_\text{d}(T=0)=1280$, $640$ and $320$ give fixed mean densities ($\rho_{\text{fixed}} = A_{\text{d}}(T=0)/L^d$) of $^{1}\rho_{\text{fixed}} = 0.2$, $^{2}\rho_{\text{fixed}} = 0.1$ and $^{3}\rho_{\text{fixed}} = 0.05$, respectively.\ \ These drops are large enough that their finite size effects are small, though still observable.\ \ Finite size effects associated with the drop size both in the Ising model and in the general discussion are  neglected here in order not to cloud the main issue of this paper.\ \ However, they can be dealt with by employing the complement approach which is discussed elsewhere \cite{moretto-04}.

In these calculations the lattice acts as the container in which a liquid drop is placed.\ \ The fluid is made up of the minority spins (e.g. down spins) with the vacuum being the majority spins (e.g. up spins).\ \ At the lowest temperatures the up spins congregate into a single cluster, the liquid drop, surrounded by vacuum.\ \ At higher temperatures the vacuum is filled with a vapor of clusters of down spins, and the liquid drop diminishes in size accordingly.\ \ Clusters in the vapor are identified via the Coniglio-Klein (CK) algorithm \cite{coniglio-80} to insure that they are physical (i.e. the cluster concentrations return Ising rather than percolation critical exponents).\ \ The largest cluster is identified as the liquid drop.\ \ It is identified geometrically (all like spin nearest neighbors bonded) in order to capture the skin thickness associated with liquid drops \cite{moretto-04}.\ \ Alternatively, the CK algorithm can be applied to the large cluster as well, but the resulting excess concentration of small clusters associated with the surface of the drop should not be used in evaluating the vapor properties.

For every temperature $T$ and magnetization $M_{\text{fixed}}$, $10^5$ thermalized realizations were generated to produce the cluster concentrations analyzed here.\ \ Temperatures are in units of spin-spin interaction strength $j$, densities are in units of spins per lattice site, molar ``volumes'' (more properly molar surfaces since the calculations are two dimensional) are given in terms of lattice sites per spin, pressure is given in units of $j$ per lattice site and cluster sizes $A$ are the number of spins in the cluster.\ \ The standard translation from Ising model to lattice gas is made through the density-magnetization relation $\rho_{\text{fixed}} = \left( 1 - M_{\text{fixed}} \right) / 2$.

The density of the vapor ${\rho}_v$ is
	\begin{equation}
	{\rho}_v  = \sum_A n_A(T) A
	\label{vapor-density}
	\end{equation}
where $n_A(T)$ is the concentration of clusters of size $A$ in the vapor.\ \ More specifically $n_A(T)$  is the yield of vapor clusters normalized to the \underline{area available to the vapor} (or volume in three dimensions) and not the total area of the lattice.\ \ The area available to the vapor is the area of the lattice $L^2$ minus the area taken up by the liquid drop: $A_\text{d}(T)+4\sqrt{A_\text{d}(T)}+n_{\text{holes}}$, where the second term accounts for the skin thickness of the liquid drop and the third term takes into account holes (bubbles) in the liquid drop.\ \ The molar volume of the vapor is $V^{v}_{m} = 1 / {\rho}_v$.\ \ The density of the liquid drop is ${\rho}_l = A_\text{d}(T) / \left( A_\text{d}(T) + n_{\text{holes}} \right)$ and the molar volume of the liquid is $V^{l}_{m} = 1 / {\rho}_l$.\ \ The exclusion of the surface ($4\sqrt{A_\text{d}(T)}$) of the drop in determining its density is in keeping with the idea that is has some skin thickness.

The pressure of the vapor $p$ is  determined by assuming, as in physical cluster models, that all interactions are exhausted by cluster  formation so that the clusters can be treated as an ideal gas  and
	\begin{equation}
	p = T \sum_A n_A(T) .
	\label{vapor-pressure}
	\end{equation}

Performing $d=2$ calculations has the benefit that the $V^{l,v}_m$ and $p$ values obtained from the cluster concentrations can be compared to the analytical result given by Lee and Yang using Onsager's solution for the $d=2$ Ising model \cite{onsager-44,lee-52.1,lee-52.2,pathria-text}.\ \ Figures~\ref{t-vm} and \ref{p-t} present such comparisons.

\begin{figure}[ht]
\includegraphics[width=8.7cm]{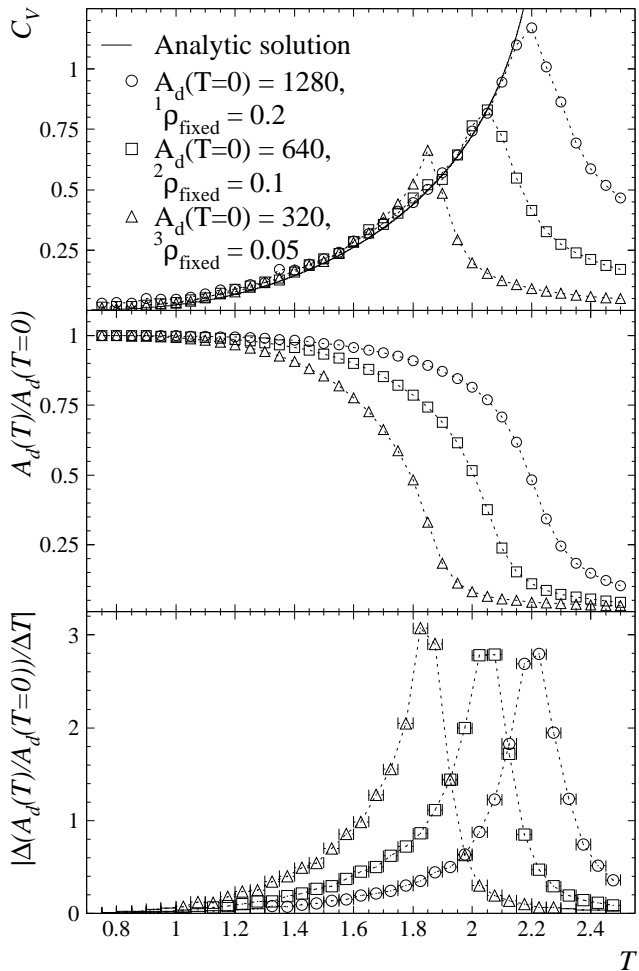}
\caption{The specific heat $C_V$ (top), liquid drop size normalized to the ground state $A_\text{d}(T)/A_\text{d}(T=0)$ (middle) and fluctuations in $A_\text{d}(T)/A_\text{d}(T=0)$ (bottom) against $T$.\ \ Peaks in $C_V$ and $\left| \Delta \left( A_\text{d}(T)/A_\text{d}(T=0)\right)/\Delta T \right|$ coincide with the deviation from coexistence.\ \ Dotted lines guide the eye.}
\label{fluc-t}
\end{figure}

\begin{figure}[ht]
\includegraphics[width=8.7cm]{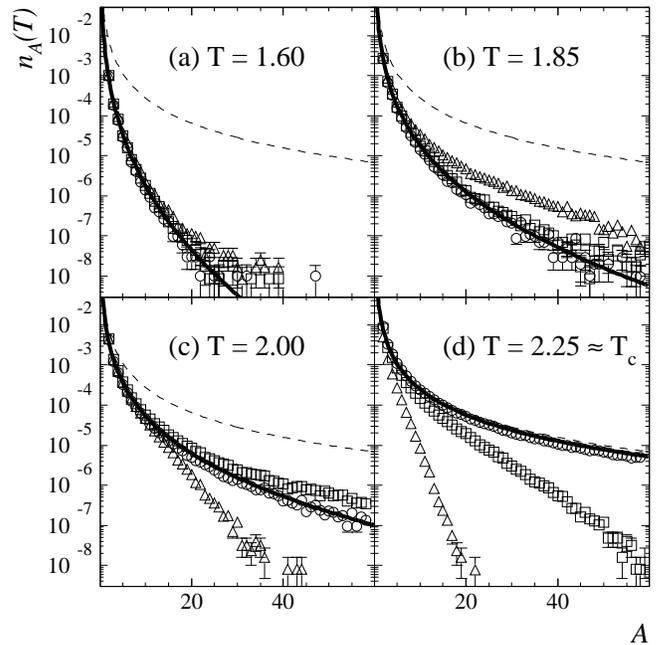}
\caption{Vapor concentrations $n_A(T)$ at different temperatures $T$ for each $\rho_{\text{fixed}}$ calculation.\ \ The curves show $n_A(T)$ (heavy solid) and $n_A(T_c)$ (light dashed)  from Eq.~(\ref{fisher-coex}).\ \ The symbol definitions are the same as in Fig.~\ref{fluc-t}.}
\label{cluster-yields}
\end{figure}

\begin{figure*}[ht]
\includegraphics[width=18.0cm]{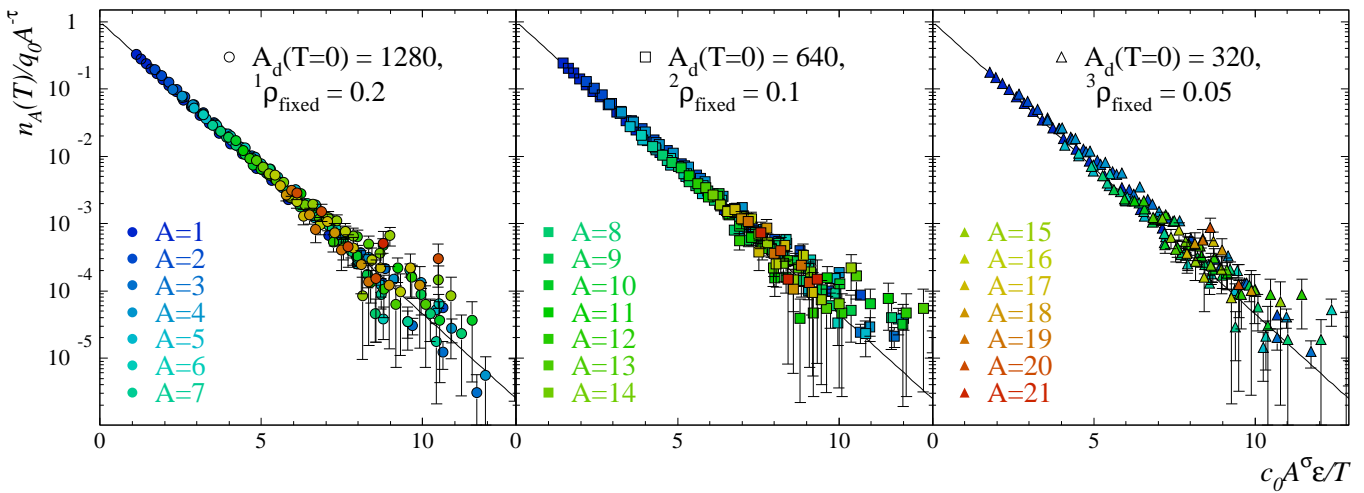}
\caption{[color online] Fisher plots: the scaled cluster concentrations fit with Eq.~(\ref{fisher-coex}) (solid line).}
\label{scalings}
\end{figure*}

Figure~\ref{t-vm} shows phase separation at low temperatures for the three $\rho_{\text{fixed}}$ calculations: there is a vapor (rightmost open symbols), with the same molar volume as that given by the analytic solution (solid curve), coexisting with a liquid drop, which has the molar volume of the liquid as given by the analytic solution.\ \ As the temperature increases, the difference between molar volumes of liquid and vapor decreases, but, because of the constraint imposed by $\rho_{\text{fixed}}$, at some point $V^{v}_{m}$ leaves the coexistence curve and attains a constant value equal to $1/\rho_{\text{fixed}}$.\ \ The smaller the ground state drop, i.e. the smaller the total number of particles in the system, the lower the temperature at which the system leaves coexistence.\ \ Deviations from the analytical solution of $V_m(T)$ before the departure from coexistence are due to the finite size of the drop, which become pronounced at large temperature where the drop is smallest.\ \ The deviations increase with increasing temperature because the drop size decreases with increasing temperature (as shown in Fig.~\ref{fluc-t}), thus increasing the drop's finite size effects.\ \ These finite size effects can be accounted for in terms of the complement \cite{moretto-04} which reduces to the Gibbs-Thomson correction \cite{krishnamachari-96} for large drops.

Figure~\ref{p-t} shows that at low temperatures all three systems agree with the analytical $T$-$p$  coexistence curve.\ \ At high temperatures all three systems develop nearly ideal gas behavior and $p \propto \rho_{\text{fixed}} T$ as shown by dashed lines.\ \ Each system leaves the coexistence line when the drop has evaporated completely; this happens at a higher temperature for the larger ground state drops.\ \ As usual, deviations from the analytical solution of $p(T)$ arise from the finite size of the drop.

The departure of the $\rho_{\text{fixed}}$ calculations from coexistence gives rise to a peak in the fluctuations of various quantities which is not related to criticality.\ \ In fact the peak in the specific heat in $\rho_{\text{fixed}}$ $d=3$ lattice gas calculations has been used to map the coexistence curve in the $\rho$-$T$ plane \cite{carmona-02}.\ \ Indeed, the top panel of Fig.~\ref{fluc-t} shows that the peak in the specific heat $C_V$  occurs just as the system leaves coexistence.\ \ The bottom panel of  Fig.~\ref{fluc-t} shows that this coincides with the peak in the fluctuations of $A_\text{d}(T)$.\ \ The peaks occur somewhat before the $\rho_{\text{fixed}}$ constraints are felt because of the effects of the finite size of the liquid drop.\ \ Evidence for this can be seen by comparing the peak locations in Fig.~\ref{fluc-t} with the departure from coexistence in Fig.~\ref{t-vm}.\ \ The coincidence is more precise as $A_\text{d}(T=0)$ increases.\ \ The peak in fluctuations observed here do not arise from criticality but because coexistence has ended.

Figure~\ref{cluster-yields} shows the vapor concentrations of the three systems and the prediction of Fisher's cluster model \cite{fisher-67} for vapor concentrations at coexistence:
	\begin{equation}
	n_A(T) = q_0 A^{-\tau} \exp \left( - \frac{c_0 A^{\sigma} \varepsilon}{T} \right)	\label{fisher-coex}
	\end{equation}
where $q_0$ is the normalization, $\tau$ is a topological critical exponent, $c_0$ is the surface energy coefficient, $\sigma$ is the perimeter to area exponent (or surface to volume in $d=3$) and $\varepsilon = (T_c - T)/T_c$.\ \ The solid curves in Fig.~\ref{cluster-yields} are obtained from Eq.~(\ref{fisher-coex}) using the $d=2$ standard parameter values given in the first column of Table~\ref{coex-t} and $c_0=8$.\ \ The dashed curves in Fig.~\ref{cluster-yields} show the critical point power law.

Figure~\ref{cluster-yields} shows the main point of this paper: the vapor cluster concentrations are well reproduced by Eq.~(\ref{fisher-coex}) when the system is at coexistence, but not when the system is away from coexistence; in other words, the location of the $\rho_{\text{fixed}}$-$T$ point in the $\rho$-$T$ plane affects the vapor concentrations.\ \ In Fig.~\ref{cluster-yields}a all three systems are at coexistence.\ \ All manifest the same vapor concentrations described by Eq.~(\ref{fisher-coex}).\ \ In Fig.~\ref{cluster-yields}b the $^{3}\rho_{\text{fixed}}$ system is departing coexistence (and entering the large fluctuation region) while $^{1}\rho_{\text{fixed}}$ and $^{2}\rho_{\text{fixed}}$ remain at coexistence.\ \ Accordingly, $^{1}\rho_{\text{fixed}}$ and $^{2}\rho_{\text{fixed}}$ manifest the same vapor concentrations described by Eq.~(\ref{fisher-coex}) while $^{3}\rho_{\text{fixed}}$ deviates above.\ \  In Fig.~\ref{cluster-yields}c, $^{3}\rho_{\text{fixed}}$ is well out of coexistence, $^{2}\rho_{\text{fixed}}$ is just leaving coexistence.\ \ Accordingly, only $^{1}\rho_{\text{fixed}}$ is still at coexistence and manifests the vapor concentration of Eq.~(\ref{fisher-coex}).\ \ In Fig.~\ref{cluster-yields}d, both $^{2}\rho_{\text{fixed}}$ and $^{3}\rho_{\text{fixed}}$  are well outside the coexistence region while $^{1}\rho_{\text{fixed}}$, leaving coexistence near $T_c$ shows a vapor concentration close to the critical point power law.

In order to further verify that the coexistence vapor concentrations are well described by Eq.~(\ref{fisher-coex}), the cluster concentration of each $\rho_{\text{fixed}}$ calculation was independently fit to Eq~(\ref{fisher-coex}) leaving $c_0$ and $T_c$ as fit parameters, using $q_0 = 1 / 2 \zeta ( \tau - 1 ) $ and fixing $\tau$ and $\sigma$ from: $( \tau - 2 ) / \sigma = \beta = 1/ 8$ and $( 3 - \tau ) / \sigma = \gamma = 7/4$.\ \ Clusters of size $7 \le A \le 21$ were included in the fit.\ \ In the thermodynamic limit the highest temperature that can be described by Eq.~(\ref{fisher-coex}) is the temperature at which the system leaves coexistence: $T \approx 2.2$, $2.05$ and $1.6$ for $^1\rho_{\text{fixed}}$,  $^2\rho_{\text{fixed}}$ and  $^3\rho_{\text{fixed}}$, respectively.\ \ However, these systems are not at the thermodynamic limit and the fluctuations grow large before those temperatures, therefore the highest temperatures fit were $T =  1.7$, $1.6$ and $1.5$ for $^1\rho_{\text{fixed}}$,  $^2\rho_{\text{fixed}}$ and  $^3\rho_{\text{fixed}}$, respectively.

\begin{figure}[ht]
\includegraphics[width=8.7cm]{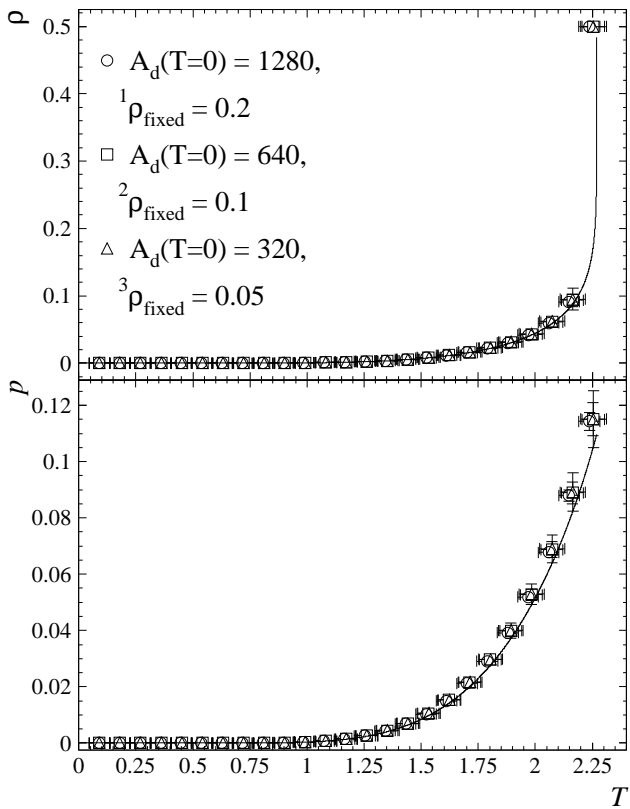}
\caption{The $T$-$\rho$ (top) and $T$-$p$ (bottom) coexistence curves.\ \ The solid curves show the analytic solution, symbols show the results of summing Eq.~(\ref{fisher-coex}) with the fit parameters in Table~\ref{coex-t}.}
\label{p-t-fisher}
\end{figure}

\begin{table}[htdp]
\caption{Parameter values and fit results}
\begin{center}
\begin{tabular}{ccccc}
\hline
        $\rho_{\text{fixed}}$     &    free     & $0.2$                       & $0.1$                        & $0.05$                        \\
        ${\chi}^{2}_{\nu}$ &          -          & $1.85$                         & $1.41$                       & $1.56$                       \\
        $c_0$                     &          -          & $8.7 \pm 0.1$             & $8.2 \pm 0.2$           & $7.8 \pm 0.2$      \\
        $T_c$                     & $2.26915$ & $2.24 \pm 0.01$        & $2.25 \pm 0.02$      & $2.25 \pm 0.03$      \\
        $p_c$                     &    $0.1097$ & $0.114\pm0.003$     & $0.115 \pm 0.002$  & $0.12 \pm 0.01$ \\
        $C_F$                    &    $0.0966$ & $0.1021\pm0.0007$& $0.102 \pm 0.001$  & $0.102 \pm 0.001$ \\
\hline
\end{tabular}
\end{center}
\label{coex-t}
\end{table}

The results are shown in Fig.~\ref{scalings} and summarized in Table~\ref{coex-t}.\ \ All systems show a good collapse to a single Fisher plot.\ \ All systems return, to within error bars, the same $T_c$ value close to the the Onsager value, also given in Table~\ref{coex-t}.\ \  This is at variance with reference \cite{gulminelli-02}, which reports changes in $T_c$ by as much as $\sim50$\% with changes in mean densities.\ \ The present results show, however, that by selecting vapor concentrations (given by a proper definition of vapor and liquid and the correct measure of the volume available to the vapor) at coexistence a consistent value of $T_c$ for all mean densities can be obtained.\ \ The change in $c_0$ observed between the systems is due to the finite size of the drop and can be accounted for by the complement approach \cite{moretto-04}.

Figure~\ref{scalings} also shows the results for clusters of size $A=1$-$6$, though they were not included in the fits.  These clusters are excluded from the fits because their perimeter (or surface in $d=3$) energies are poorly described by Fisher's ansatz: $E_\text{perimeter} = c_0 A^{\sigma}$.\ \ For example, monomers have a perimeter of $4$, and (with the $d=2$ Ising perimeter tension of $2$) $E_\text{perimeter} = 8$;  dimers have a perimeter of $6$, and $E_\text{perimeter} = 12$;  trimers have a perimeter of $8$, and $E_\text{perimeter} = 16$.\ \ Clusters of $A=4$ can have a perimeter of $8$ or $10$ and in the temperature range considered have an average perimeter of $A \sim 9$, and $E_\text{perimeter} \sim 18$.\ \ For clusters $1 \le A \le 4$ these values of $E_\text{perimeter}$ were used in place of $c_0 A^{\sigma}$ and for clusters of size $A=5$ and $6$ $E_\text{perimeter} = c_0 A^{\sigma}$ was used.

The fit results can then be used to determine the reduced phase diagram accordingly:
	\begin{equation}
	\frac{\rho}{{\rho}_c} = \frac{\sum_A A^{1-\tau}\exp\left( -\frac{c_0A^{\sigma}\varepsilon}{T}\right)}{\sum_A A^{\tau-1}}
	\label{red-den}
	\end{equation}
and
	\begin{equation}
	\frac{p}{p_c} = \frac{T\sum_A A^{-\tau}\exp\left( -\frac{c_0A^{\sigma}\varepsilon}{T}\right)}{T_c\sum_A A^{-\tau}} .
	\label{red-pres}
	\end{equation}
The sums above and those used to determine $q_0$ from $\tau$ were evaluated via methods similar to those in \cite{reuter-01}.

To determine the absolute density $\rho$ the reduced density  from Eq.~(\ref{red-den}) was multiplied by the known critical density of the system $\rho_c = 1/2$. The top plot in Fig.~\ref{p-t-fisher} shows the absolute vapor density at coexistence as a function of temperature.\ \ To obtain the bottom plot of Fig.~\ref{p-t-fisher}, the $T$-$p$ coexistence curve, the values of $T_c$ and ${\rho}_c$ are used with the compressibility factor $C_f = p_c / {\rho}_c T_c = \sum_A n_A(T) / \sum_A A n_A(T)$.\ \ All systems return the same coexistence curve and agree with the analytical solution to within error bars.

In conclusion, it has been shown that: in agreement with thermodynamic expectations, the properties of the $d=2$ lattice gas (Ising) vapor at coexistence are independent of mean particle number density; their Fisher plots are unique; and lead to unique critical temperatures and phase diagrams.

\noindent {\bf Acknowledgments}

This work was supported by the US Department of Energy.\ \ The authors thank K. A. Bugaev for his input.

\end{document}